\documentclass[aps,onecolumn]{revtex4}
\usepackage{graphicx}
\usepackage{subfigure}
\usepackage{caption2}
\usepackage{float}
\usepackage{amsmath}
\usepackage{amssymb}

\input{tcilatex}

\begin{document}

\title{All-optical switching in a two-channel waveguide with cubic-quintic
nonlinearity}
\author{Rodislav Driben$^{1}$, Boris A. Malomed$^{1}$, and Pak L. Chu$^{2}$}

\begin{abstract}
We consider dynamics of spatial beams in a dual-channel waveguide with
competing cubic and quintic (CQ) nonlinearities. Gradually increasing the
power in the input channel, we identify four different regimes of the
pulse's coupling into the cross channel, which alternate three times between
full pass and full stop, thus suggesting three realizations of switching
between the channels. As in the case of the Kerr (solely cubic)
nonlinearity, the first two regimes are the linear one, and one dominated by
the self-focusing nonlinearity, with the beam which, respectively,
periodically couples between the channels, or stays in the input channel.
Further increase of the power reveals two novel transmission regimes, one
characterized by balance between the competing nonlinearities, which again
allows full coupling between the channels, and a final regime dominated by
the self-defocusing quintic nonlinearity. In the latter case, the situation
resembles that known for a self-repulsive Bose-Einstein condensate trapped
in a double-well potential, which is characterized by strong symmetry
breaking; accordingly, the beam again abides in the input channel, contrary
to an intuitive expectation that the self-defocusing nonlinearity would push
it into the cross channel. The numerical results are qualitatively explained
by a simple analytical model based on the variational approximation.
\end{abstract}

\maketitle

\address{$^{1}$Department of Interdisciplinary Studies, School of Electrical
Engineering,\\
Faculty of Engineering, Tel Aviv University, Tel Aviv 69978, Israel\\
$^{2}$Optoelectronics Research Centre, Department of Electronics Engineering,\\
City University of Hong Kong, Tat Chee Avenue, Kowloon, Hong Kong}

\section{Introduction and the model}

\label{intro}

Dynamics of spatial light beams in nonlinear planar waveguides with a
multi-channel built-in structure is a subject of fundamental interest to
nonlinear optics, and has vast potential for applications to all-optical
routing of data streams. The basic model of the multi-channel system amounts
to the nonlinear Schr\"{o}dinger (NLS) equation for the local amplitude $%
\Psi (z,x)$ of the electromagnetic wave, where $z$ and $x$ are the
propagation and transverse coordinates in the waveguide. In a normalized
form, the NLS equation is \cite{Wang} 
\begin{equation}
i\Psi _{z}+\Psi _{xx}-U(x)\Psi +\delta n\left( |\Psi |^{2}\right) \Psi =0,
\end{equation}%
where the second term accounts for the diffraction of light in the paraxial
approximation, $U(x)$ is the profile of the transverse modulation of the
refractive index that accounts for the multi-channel structure, and $\delta
n(|\Psi |^{2})$ is a nonlinear correction to the refractive index. In the
case of the ordinary self-focusing (SF)\ Kerr nonlinearity, $\delta n\left(
|\Psi |^{2}\right) =n_{2}|\Psi |^{2}$, with $n_{2}>0$.

The interplay of the diffraction and SF nonlinearity gives rise to
channel-trapped spatial solitons. In the model with the Kerr nonlinear term
and harmonic transverse modulation, $U(x)=U_{0}\cos \left( 2\pi x/L\right) $%
, trapped solitons and a possibility of their switching were studied in Ref. 
\cite{Wang}. It was found that the soliton's integral power, $%
P=\int_{-\infty }^{+\infty }|\Psi (x)|^{2}dx,$which is a dynamical invariant
of the NLS equation, may take any value, $0<P<\infty $, the entire solution
family being stable. Switching, i.e., transfer of the soliton into an
adjacent channel, was demonstrated under the action of a ``hot spot" modeled
by an extra term $\sim \delta (x-x_{0})\delta (z-z_{0})\Psi $ in the
equation, with $x_{0}$ fixed at the midpoint between the channels.

The same model as in Ref. \cite{Wang}, with $z$ replaced by time $t$ (but
without the hot-spot term) was later introduced for an effectively
one-dimensional Bose-Einstein condensate (BEC) with attraction between
atoms, loaded in an optical-lattice potential \cite{KonotopEPL}. Spatial
solitons in a model of the Kerr-nonlinear optical waveguide with the
periodic transverse modulation in the form of the \textit{Kronig-Penney}
profile (a periodic array of rectangular troughs, see below) were considered
in Refs. \cite{Smerzi}. In optics, a system of parallel guiding cores can be
created not only through a small periodic variation of the material's
refractive index, but also in a \textit{virtual form}, by means of a system
of strong transverse laser beams illuminating a sample of a photorefractive
material in the ordinary polarization, while the signal beam is launched in
the extraordinary polarization \cite{Moti}.

The functionality of optical beam-handling schemes may be greatly enhanced
in media with competing nonlinearities, the simplest example of which is
provided by a combination of the SF cubic and self-defocusing (SDF) quintic
terms. Observations of the cubic-quintic (CQ) optical nonlinearity were
reported in the PTS crystal \cite{PTS}, chalcogenide glasses \cite{glass}
and some other materials \cite{organic}. In fact, the CQ nonlinearity occurs
due to an intrinsic nonlinear resonance in the material, which also gives
rise to strong two-photon absorption \cite{Leblond}; however, experiments
with spatial solitons require short propagation distances, on the order of a
few centimeters \cite{spatial-soliton,Moti}, over which effects of the loss
may be insignificant \cite{YiFan}.

The NLS equation with the CQ nonlinearity and the multi-channel guiding
structure can be rescaled into the following form \cite{we}: 
\begin{equation}
i\Psi _{z}+\Psi _{xx}=U(x)\Psi -2|\Psi |^{2}\Psi +|\Psi |^{4}\Psi .
\label{eq1}
\end{equation}
In the uniform medium [$U(x)=0$], Eq. (\ref{eq1}) gives rise to a well-known
family of stable solitons \cite{Canada}, 
\begin{equation}
\Psi (x,z)=\exp \left( ikz\right) \sqrt{\frac{2k}{1+\sqrt{1-4k/3}\cosh
\left( 2\sqrt{k}x\right) }},
\end{equation}
with the propagation constant $k$ taking values in a finite interval, $%
0<k<3/4$. In Ref. \cite{we}, solitons were studied in CQ model (\ref{eq1})
with a single guiding channel of a rectangular shape. A remarkable feature
of the channel-trapped CQ solitons is \emph{bistability} of the
guided-soliton family: in the same region as mentioned above, $0<k<3/4 $,
the channel supports a single soliton state, but the full existence region
extends to an additional interval, $3/4<k<k_{\max }$, where two different
solitons are found for each $k$, \emph{both} being stable (the soliton
bistability does not occur in the same guiding channel if the model contains
only the cubic nonlinearity \cite{gh}). The CQ nonlinearity was combined
with the multi-channel structure of the Kronig-Penney type in Ref. \cite%
{we-all}. In addition to fundamental solitons, many families of higher-order
solitons, also featuring the bistability, were found in the latter model.

For applications to all-optical switching, more relevant solutions are not
static solitons, but rather dynamical states featuring periodic transfer (%
\textit{coupling}) of the beam between two adjacent guiding cores. In this
context, the most appropriate model is one with two channels, similar to
nonlinear directional couplers operating in the temporal domain, which are
based on dual-core fibers (see Refs. \cite{Jena} and references therein).
The objective of the present work is to report numerical results and some
analytical approximations for the beam switching in the dual-channel CQ
model, which is based on Eq. (\ref{eq1}) with 
\begin{equation}
U(x)=\left\{ 
\begin{array}{ll}
0, & |x|~<\frac{1}{2}L~\mathrm{and~}|x|~>\frac{1}{2}L+D, \\ 
-U_{0}, & \frac{L}{2}<|x|~<\frac{1}{2}L+D,%
\end{array}%
\right.  \label{DC}
\end{equation}
where $D$, $U_{0}$ and $L$ are, respectively, the width and depth of each
channel, and the thickness of the buffer layer between them. A fundamental
difference from earlier studied two-channel models with the Kerr
nonlinearity is expected dependence of the coupling on the beam's input
power $P$: in the Kerr model, the increase of $P$ leads simply to transition
from the nearly-linear-coupling regime to a nonlinear one, in which the beam
stays in the input channel (coupling suppression). In the CQ model, a more
sophisticated dependence on $P$ may be anticipated, due to the competition
between the cubic SF and quintic SDF terms. The main qualitative result
demonstrated below is that the coupling suppression is changed by
restoration of the coupling at larger values of $P$ (at which the SF and SDF
terms are in balance), and finally again by suppression of the coupling at
very large $P$, when the quintic term dominates, even if the suppression of
coupling by the self-\emph{defocusing} nonlinearity is a counterintuitive
effect. In Section \ref{numerical}, we display typical examples of the four
distinct transmission modes, and then present a diagram which provides for
overall description of the switching in the dual-channel waveguide with the
CQ nonlinearity. Using values of strengths of the cubic and quintic SF and
SDF nonlinearities in available materials, we estimate that the switching
can be achieved at power densities of the input beam on the order of GW/cm$%
^{2}$, the corresponding coupling distance being $_{\sim }^{<}~1$ cm.

In addition to the numerical results, in Section \ref{analytical} we report
results of a simple analytical approximation that provide for a qualitative
explanation of the switchings in the present model. The analytical approach
is based on a variational approximation, similar to that developed earlier
in detail for temporal signals (solitons) in dual-core fibers \cite%
{earlyVA,weJOSAB,Jena} (see also Chapter 6 of review \cite{Progress}).

\section{Coupling dynamics and switching diagrams}

\label{numerical}

To simulate the transmission of the spatial beam in the system described by
Eq. (\ref{eq1}), we integrated this equation with initial conditions (at $%
z=0 $) corresponding to a Gaussian beam, with amplitude $A$ and width $W$,
launched into one channel:

\begin{equation}
\Psi _{0}=A\exp \left[ -\frac{1}{W^{2}}\left( x-\frac{L+D}{2}\right) ^{2}%
\right]  \label{input}
\end{equation}
[$x=\left( L+D\right) /2$ is the center of the first channel as per Eq. (\ref%
{DC})]. The splitting of light between the two channels is quantified by
respective integral powers,

\begin{equation}
P_{1}=\int\limits_{L/2}^{\left( L+D\right) /2}\left\vert \Psi \right\vert
^{2}dx,~P_{2}=\int\limits_{-\left( L+D\right) /2}^{-L/2}\left\vert \Psi
\right\vert ^{2}dx~.  \label{Power}
\end{equation}
Equation (\ref{eq1}) was solved numerically by means of the split-step
method. The input power was gradually increased, and the evolution of the
power distribution across the waveguide was monitored in the course of the
simulations. Collecting numerical results, it was found that they are
adequately represented by fixing the parameters at $L=D=U_{0}=W=2$. This
implies that the width of the channels is equal to the thickness of the
buffer layer between them, and the diffraction and channel-trapping terms
are comparable to the nonlinear ones when they are in balance. Results for
other values of the parameters will be displayed too.

If the input power is small, nonlinear terms in Eq. (\ref{eq1}) are
negligible in comparison to the diffraction, which leads to the ordinary
picture of the periodic coupling of power between the channels, as shown in
Fig. 1. Note that the transmission length equal to coupling length $Z_{%
\mathrm{coupl}}$, i.e., half-period of the coupling (as in Fig. 1), provides
for the full transfer of the beam into the cross channel.

Increase of the input power makes the SF cubic nonlinearity dominant,
switching (as usual) the transmission mode from the periodic coupling to one
with the power abiding in the input channel. This situation is illustration
by Fig. 2.

The two transmission modes shown in Figs. 1 and 2, and the switching between
them are qualitatively the same as in ordinary dual-core models with the
Kerr nonlinearity \cite{Jena}. Switching to an essentially new transmission
regime, which is specific to the CQ medium, occurs with further increase of
the power, in a range where the SF cubic and SDF quintic terms balance each
other, rendering the situation similar to that in the linear model. This new 
\textit{quasi-linear} regime is illustrated by Fig. 3. It is noteworthy that
the linear-coupling mode in Fig. 1, and its strongly nonlinear but similar
counterpart in Fig. 3 have nearly equal coupling lengths: for equal values
of the system's parameters and beam's width, $Z_{\mathrm{coupl}}^{\mathrm{%
(nonlin)}}/Z_{\mathrm{coupl}}^{\mathrm{(lin)}} \approx 7/6$.

Extra (final) switching occurs, if one keeps increasing the input power, to
a transmission regime in which the SDF quintic term dominates. ``Naively",
one might expect that SDF nonlinearity would only enhance coupling between
the guiding cores; however, it actually \emph{suppresses }the coupling, as
shown in a typical example in Fig. 4. As observed in the figure, this regime
is not completely equivalent to one in which the coupling was suppressed by
the dominant SF cubic term (cf. Fig. 2), as the present transmission mode
features conspicuous oscillations; nevertheless, the coupling is largely
suppressed in this case too.

A result similar to that demonstrated in Fig. 4 is known in a model of a
Bose-Einstein condensate (BEC) with repulsion between atoms (which is
formally tantamount to SDF cubic nonlinearity) trapped in a double-well
potential, the two potential minima playing the same role as the
dual-channel configuration. The corresponding Gross-Pitaevskii equation (the
counterpart of the NLS equation for the BEC) gives rise to stable \emph{\
asymmetric} states (see Ref\cite{BECtheory} and references therein). This
symmetry-breaking effect has also been observed in experiment \cite%
{BECexperiment} [in Ref. \cite{Zhigang}, a similar effect was demonstrated,
theoretically and experimentally (using a photorefractive crystal), in a
medium with saturable nonlinearity].

Results of systematic simulations are summarized in the form of a \textit{%
switching diagram}, which is displayed in Fig. 5. To measure the coupling
rate, we define $\varepsilon \equiv \left( P_{2}\right) _{\max }/P$, as the
maximum share of the total input power ($P$), which is coupled into the
second channel. The diagram plots $\varepsilon $ versus the total power
itself; to demonstrate the generic character of the dependence, it is
presented for three different values of the width of the input pulse (\ref%
{input}) (including $W=2$, for which examples of the transmission modes are
displayed in Figs. 1-4). The regimes with $\varepsilon $ approaching $1$,
and with $\varepsilon \ll 1$ are, respectively, those featuring nearly
complete periodic transfer of the power between the channels, and
suppression of the coupling. In accordance with what was displayed above by
dint of examples, the strong coupling takes place at small $P$, which
corresponds to the nearly linear transmission, and in a relatively narrow
interval of $P$ providing for the balance between the SF and SDF
nonlinearities. In two broad intervals, where either the cubic term or its
quintic competitor dominate, the coupling is suppressed. For applications, a
beneficial feature of the $\varepsilon (P)$ dependences in Fig. 5 is the
steep transition between regimes of the two types, i.e., \emph{sharp
switching}. Figure 5 also suggests that exact values of the power at which
the switchings take place can be tuned by adjusting the width of the input
pulse.

To conclude this section, we note that, with typical values of the cubic and
quintic nonlinear coefficients for chalcogenide glasses reported in Refs. 
\cite{Leblond}, $n_{2}=2.2\times 10^{-4}$ cm$^{2}$/GW, and $n_{4}=-7.9\times
10^{-5}$ cm$^{4}$/(GW)$^{2}$ (they are defined so that the total nonlinear
correction to the refractive index is $\delta n=n_{2}I+n_{4}I^{2}$, where $I$
is the power density in the optical beam), the switchings specific to the CQ
medium are attained in a ballpark of $I=3$ GW/cm$^{2}$, which provides for
the balance between the SF and SDF effects, and is achievable in the
experiment \cite{glass}. Analysis of the corresponding wave-propagation
equation written in physical units demonstrates that the respective coupling
length (in the quasi-linear regime of balanced SF and SDF) may take values
in the range of $0.1$ to $1$ cm.

\section{Analytical approximation}

\label{analytical}

The underlying equation (\ref{eq1}) can be derived from the Lagrangian 
\begin{equation}
L=\int_{-\infty }^{+\infty }\left[ \frac{i}{2}\left( \Psi ^{\ast }\Psi
_{z}-\Psi _{z}^{\ast }\Psi \right) -\left\vert \Psi _{x}\right\vert
^{2}-U(x)|\Psi |^{2}+|\Psi |^{4}-\frac{1}{3}|\Psi |^{6}\right] dx.  \label{L}
\end{equation}%
To apply the variational approximation to the present model, we assume a
situation with very narrow and deep channels, separated by a broad buffer
layer. The field configuration is then approximated as follows: inside each
channel, i.e., in regions $\left\vert x-\left( L\pm D\right) /2\right\vert
<D/2$ [see Eq. (\ref{DC})], 
\begin{equation}
\Psi (x,z)=A_{\pm }(z)\cos \left( \pi \frac{x\mp \left( L+D\right) /2}{D}%
\right)   \label{inner}
\end{equation}%
with complex amplitudes $A_{\pm }$ (this ansatz emulates the ground state of
a quantum-mechanical particle in an infinitely deep potential box, therefore
it vanishes at edges of the channel). The form of the ansatz outside the
channels is also borrowed from quantum mechanics, emulating a superposition
of exponentially decaying ground-state wave functions behind the edges of
very deep but finite potential boxes, 
\begin{equation}
\Psi (x,z)=\sum_{+,-}A_{\pm }(z)\exp \left( -\sqrt{U_{0}-\left( \pi
/D\right) ^{2}}\left\vert x\mp \frac{L+D}{2}\right\vert \right) .
\label{outer}
\end{equation}%
The substitution of the inner and outer parts of the ansatz, (\ref{inner})
and (\ref{outer}), in Lagrangian (\ref{L}) and integration lead to the
following effective Lagrangian: 
\begin{eqnarray}
\frac{4}{D}L_{\mathrm{eff}} &=&\sum_{+,-}\left\{ i\left[ A_{\pm }^{\ast
}A_{\pm }^{\prime }-A_{\pm }\left( A_{\pm }^{\ast }\right) ^{\prime }\right]
+2\left[ U_{0}-\left( \frac{\pi }{D}\right) ^{2}\right] \left\vert A_{\pm
}\right\vert ^{2}+\frac{3}{2}\left\vert A_{\pm }\right\vert ^{4}-\frac{5}{12}%
\left\vert A_{\pm }\right\vert ^{6}\right\}   \notag \\
&&+\frac{4}{D}\sqrt{U_{0}-\left( \frac{\pi }{D}\right) ^{2}}\exp \left( -%
\sqrt{U_{0}-\left( \frac{\pi }{D}\right) ^{2}}\left( L+D\right) \right)
\left( A_{+}A_{-}^{\ast }+A_{-}A_{+}^{\ast }\right) ,  \label{Leff}
\end{eqnarray}%
where the prime stands for $d/dz$. Note that the last term in (\ref{Leff}),
which accounts for the interaction between the beams trapped in the two
channels, was calculated by means of a general method elaborated in Ref. 
\cite{me} (see also section 2.3 in review \cite{Progress}). This method only
assumes that, in a vicinity of each channel, the tail created by the other
beam is much weaker than the field of the beam trapped in the given channel
(note that $L+D$ is the distance between centers of the two channels).

The Euler-Lagrange equations, $\partial L_{\mathrm{eff}}/\partial A_{\pm
}^{\ast }-\left[ \partial L_{\mathrm{eff}}/\partial \left( A_{\pm }^{\ast
}\right) ^{\prime }\right] ^{\prime }=0$ can be cast in a real form, by
defining $A_{\pm }\equiv R_{\pm }e^{i\phi _{\pm }}$, where $R_{\pm }$ and $%
\phi _{\pm }$ are real amplitudes and phases: 
\begin{eqnarray}
R_{+}^{\prime } &=&\kappa R_{-}\sin \phi ,~R_{-}^{\prime }=-\kappa R_{+}\sin
\phi ,  \label{RR} \\
\phi _{\pm }^{\prime } &=&\left[ -\frac{3}{2}+\frac{5}{8}\left(
R_{+}^{2}+R_{-}^{2}\right) \right] \left( R_{\mp }\right) ^{2}+\kappa \left(
\cos \phi \right) \frac{R_{\mp }}{R_{\pm }},  \label{phiphi}
\end{eqnarray}
where the effective constant of the tunnel coupling between the two guiding
cores is 
\begin{equation}
\kappa \equiv \frac{4}{D}\sqrt{U_{0}-\left( \frac{\pi }{D}\right) ^{2}}\exp
\left( -\sqrt{U_{0}-\left( \frac{\pi }{D}\right) ^{2}}\left( L+D\right)
\right) .  \label{kappa}
\end{equation}
An obvious consequence of Eqs. (\ref{RR}) and (\ref{phiphi}) is the
conservation of $P\equiv R_{+}^{2}+R_{-}^{2}$ (it is proportional to the
total power of the beam), and the fact that an equation for $\phi _{+}+\phi
_{-}$ is detached from the other equations. Thus, defining 
\begin{equation}
R_{+}\equiv \sqrt{P}\cos (\theta /2),R_{-}\equiv \sqrt{P}\sin (\theta
/2),~\phi \equiv \phi _{+}-\phi _{-},  \label{theta}
\end{equation}
\begin{equation}
\lambda \equiv (P/8)\left( 12-5P\right) ,  \label{lambda}
\end{equation}
we end up with a dynamical system 
\begin{equation}
\begin{array}{c}
\theta ^{\prime }=-\kappa \sin \phi , \\ 
\phi ^{\prime }=\lambda \cos \theta -\kappa \left( \cos \phi \right) \cot
\theta ,%
\end{array}
\label{final}
\end{equation}
which conserves the Hamiltonian, 
\begin{equation}
H=\left( \kappa \cos \phi -\frac{1}{2}\lambda \sin \theta \right) \sin
\theta .  \label{H}
\end{equation}

Dynamical system (\ref{final}) is similar to that which was derived, in the
simplest version of the variational approximation, for a two-component
temporal soliton in a dual-core optical fiber with the Kerr nonlinearity 
\cite{earlyVA,weJOSAB}. The only significant difference is that the present
CQ model gives rise to nonlinearity coefficient $\lambda $ which changes its
sign with the increase of the beam's power $P$, see Eq. (\ref{lambda}) [$%
\kappa $ is always positive, according to its definition (\ref{kappa})].

In the framework of this approximation, solutions corresponding to various
coupling modes found above by means of numerical simulations of Eq. (\ref%
{eq1}) with initial condition (\ref{input}) are obtained from the initial
condition with the beam launched into one channel, i.e., $\theta =0$ or $%
\theta =\pi $, according to Eqs. (\ref{theta}). As seen from Eq. (\ref{H}),
this initial condition yields $H=0$, hence the dynamical trajectory in the $%
\left( \theta ,\phi \right) $ plane, $H=\mathrm{const}$, is generated by the
equation 
\begin{equation}
\cos \phi =\frac{\lambda }{2\kappa }\sin \theta .  \label{trajectory}
\end{equation}

It follows from Eq. (\ref{trajectory}) that the trajectories take completely
different forms in cases $|\lambda |<2\kappa $ and $|\lambda |>2\kappa $,
see examples shown in Figs. 6(a) and (b). The former case corresponds to
weak effective nonlinearity, i.e., to the coupling mode with either small
input power, or with balanced SF and SDF terms, cf. Figs. 1 and 3,
respectively. In qualitative agreement with the results of numerical
simulations of Eq. (\ref{eq1}), in this (quasi-linear) case the trajectories
run along the $\theta $ direction, which means periodic transfer of power
between the channels. Note that Eq. (\ref{lambda}) makes it possible to
predict intervals of power $P$ in which the quasi-linear coupling occurs: $%
P<(2/5)\left( 3-\sqrt{9-20\kappa }\right) $, and $(2/5)\left( 3+\sqrt{%
9-20\kappa }\right) <P<(2/5)\left( 3+\sqrt{9+20\kappa }\right) $ [these
expressions imply constraint $\kappa <9/20$, which complies with the fact
that expression (\ref{kappa}) yields small $\kappa $]. On the other hand, in
the case of strong nonlinearity, $|\lambda |>2\kappa $, i.e., in intervals $%
(2/5)\left( 3-\sqrt{9-20\kappa }\right) <P<(2/5)\left( 3+\sqrt{9-20\kappa }%
\right) $ and $P>(2/5)\left( 3+\sqrt{9+20\kappa }\right) $, Fig. 6(b)
displays trajectories confined to a vicinity of $\theta =0$ or $\theta =\pi $%
, which corresponds to the suppression of coupling, cf. Figs. 2 and 4. Thus,
the variational approximation provides for a qualitatively correct
explanation of the numerical results presented above.

\section{Conclusion}

\label{conclusion}

We have investigated possibilities of realizing power-controlled switching
of spatial light beams in the dual-channel planar waveguide with the
cubic-quintic nonlinearity. The interplay of the transverse diffraction and
self-focusing (cubic) and defocusing (quintic) nonlinearities gives rise to
two transmission modes similar to those known in models with the cubic
nonlinearity, i.e., periodic coupling in the nearly linear situation, and
suppression of the coupling in the mode dominated by the cubic term, and two
additional regimes, one with the periodic coupling restored due to the
balance between the competing nonlinearities, and the other (final)
transmission mode with the coupling suppressed again when the
self-defocusing quintic term dominates. Three switchings (between the four
distinct transmission regimes) are not strongly affected by variation of the
system's parameters, but they feature sharpness with variation of the power,
especially the reverse switching from the coupling-suppressed mode dominated
by the cubic nonlinearity to the strong-coupling one provided by the balance
between the self-focusing and defocusing nonlinearities. A simple analytical
approximation, which reduces the model to a Hamiltonian dynamical system
with one degree of freedom, qualitatively explains all these coupling models
and switchings between them.

This work was supported, in a part, by a Strategic Research Grant of the
City University of Hong Kong (project No.7001705), and by the Israel Science
Foundation through a Center-of-Excellence grant No. 8006/03. R.D. and B.A.M.
appreciate hospitality of the Optoelectronics Research Centre at the
Department of Electronics Engineering, City University of Hong Kong.

\newpage

\section*{Figure Captions}

Fig. 1. (Color online) The first half-period of the quasi-linear coupling
regime, for input beam (\ref{input}) with $A=0.2$ and $W=2$: (a) contour
plots of the power distribution; (b) integral powers in the channels,
defined as per Eq. ( \ref{Power}), vs. the propagation distance.

Fig. 2. (Color online) The same as in Fig. 1, but after the switch to the
transmission regime dominated by the self-focusing cubic term. In this case,
the input amplitude is $A=1$, i.e., the total power is larger by a factor of 
$25$ than in the linear regime displayed in Fig. 1. A longer transmission
distance than in Fig. 1 is shown here, to make it sure that the coupling to
the the cross channel never commences.

Fig. 3. (Color online) The same as in Fig. 2, but after the \textit{reverse
switch} to the regime in which the self-focusing and defocusing terms (cubic
and quintic ones) are in balance, thus again allowing strong coupling
between the channels. In this case, the input amplitude in Eq. (\ref{input})
is $A=\sqrt{2}$, i.e., the total power is twice as large as in the case
shown in Fig. 2. Unlike Fig. 1, about $1.5$ periods of the coupling is shown
here, to demonstrate the stability of the periodic coupling (actually, it
remains stable indefinitely long, in more extensive simulations).

Fig. 4. (Color online) The same as Fig. 3, but after the final switch to the
transmission mode with suppressed coupling. The amplitude of the input pulse
is $A=1.8$, hence the total input power exceeds that in Fig. 3 by a factor
of $1.\,\allowbreak 62$.

Fig. 5. (Color online) Switching diagrams, which show the maximum share of
the input power coupled into the cross channel, as a function of the input
power itself. The diagrams are presented for initial Gaussian beams (\ref%
{input}) with three different values of the width, $W=1$, $2$, and $2\sqrt{2}
$ (recall examples of the four transmission regimes displayed in Figs. 1-4
pertain to $W=2$).

Fig. 6. Typical dynamical trajectories in the phase plane $\left( \theta
,\phi \right) $, as generated by Eq. (\ref{trajectory}): (a) $\lambda
/(2\kappa )=0.5$, which corresponds to the quasi-linear coupling mode; (b) $%
\lambda /(2\kappa )=2$, corresponding to the coupling-suppressed mode,
dominated by the nonlinearity.

\end{document}